\documentstyle[prd,aps,epsfig]{revtex}
\begin{document}
\draft

\title{Partial widths for the decays $\eta(1295)\to\gamma\gamma$
and $\eta(1440)\to\gamma\gamma$}
\author{A. V. Anisovich, V. V. Anisovich, V. A. Nikonov}
\address{St.Petersburg Nuclear Physics Institute, Gatchina, 188350,
Russia}
\author{L. Montanet}
\address{CERN, CH-1211 Geneva 23, Switzerland}
\maketitle

\begin{abstract}
We  discuss $\gamma\gamma$ partial widths
of pseudoscalar/isoscalar mesons $\eta(M) $ in the mass region
$M\sim 1000-1500$ MeV. The transition amplitudes
$\eta(1295)\to\gamma\gamma$ and
$\eta(1440)\to\gamma\gamma$
are studied within an assumption that the decaying
mesons are the members of the first radial excitation nonet
$2^1S_0q\bar q$. The calculations show that
partial widths being of the order of 0.1 keV are
dominantly due to the $n\bar n$  meson component while the
contribution of the $s\bar s$ component is small.
\end{abstract}

\pacs{14.40.Aq, 13.40.Fn, 13.40.Hq}

The two--photon decays of the pseudoscalar mesons served a great deal
 of information on the structure of the basic $1^1S_0q\bar q$  nonet.
The value of the partial width $\pi^0\to\gamma\gamma$ gave one of the
first experimental evidence for colour structure of quarks. The decays
$\eta\to\gamma\gamma$ and $\eta'\to\gamma\gamma$ provide an information
on the quark/gluon content of these mesons.

The partial $\gamma\gamma$-widths  of pseudoscalar mesons
belonging to the
basic nonet $1^1S_0q\bar q$ are relatively large:
$\Gamma_{\pi^0\to\gamma\gamma}=7.2\pm 0.5$ eV,
$\Gamma_{\eta\to\gamma\gamma}=0.46\pm 0.04$ keV,
$\Gamma_{\eta'\to\gamma\gamma}=4.27\pm 0.19$ keV \cite{PDG} thus giving
a possibility to mesure not only partial widths
but  transition form factors $\pi^0\to\gamma^*(Q^2)\gamma$,
$\eta\to\gamma^*(Q^2)\gamma$, $\eta'\to\gamma^*(Q^2)\gamma$ over a broad
range of photon virtualities, $Q^2\le 20$ GeV$^2$ \cite{exper}.
These data made it possible \cite{AMN}:\\
(i) to restore the wave functions of $\eta$ and $\eta'$ (for both
$n\bar n$ and $s\bar s$ components),\\
(ii) to estimate the gluonium admixture in $\eta$ and
$\eta'$,\\
(iii) to restore the vertex function for the transition $\gamma\to
q\bar q$ (or photon wave function) as a function of the $q\bar q$
invariant mass.

The same method as used in \cite{AMN} for the analysis of the
basic pseudoscalar mesons can be applied for a study
of $\gamma\gamma$ decays  of
the first radial excitation mesons: in the present paper
we discuss these processes.

The search for exotics in the pseudoscalar/isoscalar sector ultimately
requires the investigation of $\eta$-mesons of the $2^1S_0
q\bar q$ nonet: in the framework of this investigation program, here we
calculate
partial widths $\eta(1295)\to\gamma\gamma$ and
$\eta(1440)\to\gamma\gamma$ under the assumption
that the mesons $\eta(1295)$ and $\eta(1440)$ are
members of the $2^1S_0q\bar q$ nonet. The state $\eta(1440)$ (old name
is $E(1407)$ \cite{baillon}) attracts our special attention: it is
considered during long time as a state with possible rich gluonic
component.

{\bf Transition form factors and partial widths.}
A partial width for the decay $\eta(M)\to\gamma\gamma$ is determined as
$\Gamma_{\eta (M)\to\gamma\gamma}=
\pi\alpha^2 M^3 F^2_{\eta(M)\to\gamma\gamma}(0)/4 $,
where $M(M)$ is the mass of the $\eta$-meson, $\alpha=1/137$, and
$F_{\eta(M)\to\gamma\gamma}(0)$ is the form factor of the considered
decay.

In Ref. \cite{AMN},
the form factor $F_{\eta\to\gamma^*\gamma}(Q^2) $
was calculated for the virtual
photon $\gamma^*(Q^2)$; the decay form factor is given by the limit
$Q^2\to 0$. The decay form factor $
F_{\eta(M)\to\gamma\gamma}(0)$ reads \cite{AMN}:
\begin{eqnarray}
F_{\eta(M)\to\gamma\gamma}(0)&=&
\frac 1{6\sqrt 3 \pi^3}\int\frac{dxd^2{k_\perp}}{x(1-x)^2}
\left[\frac {5m}{\sqrt 2}\cos{\phi}\;\Psi_{n\bar n}(s)
                             \Psi_{\gamma\to n\bar n}(s)
     +m_s\sin{\phi}\;\Psi_{s\bar s}(s)
                             \Psi_{\gamma\to s\bar s}(s)\right].
\end{eqnarray}
Two terms in the square brackets refer to $n\bar n$ and $s\bar s$
components of the $\eta(M)$-meson.
The flavour wave function is determined as
$\psi_\eta (M)=
 \cos{\phi}\;n\bar n +\sin{\phi}\;s\bar s $
where $\phi$ is mixing angle and $n\bar n=(u\bar u+d\bar d)/\sqrt 2$;
$m$ and $m_s$ are masses of the non-strange  and  strange  constituent
quarks.

The wave functions for
$n\bar n$ and $s\bar s$ components are written as
$\Psi_{n\bar n}(s)$ and $\Psi_{s\bar s}(s)$
where $s$ is $q\bar q$ invariant mass squared. In terms of the light
cone variables $(x, \vec k_{\perp})$,
the $q\bar q$ invariant mass reads
$s=(m^2+k_\perp^2 )/x(1-x)$.
The photon wave function $\Psi_{\gamma\to q\bar q}(s)$ was found in
Ref. \cite{AMN}: it is shown in Fig. 1a.

{\bf Wave functions of $\eta(M)$-mesons}.
We approximate wave functions of the $\eta(M)$-mesons in the
one--parameter exponential form. For the basic multiplet and
first radial excitation nonet, the wave functions
are determined as follows:
\begin{equation}\label
{wfparam}
\Psi^{(0)}_\eta(s)=Ce^{-bs}, \quad
\Psi^{(1)}_\eta(s)=C_1(D_1 s-1)e^{-b_1 s}.
\end{equation}
The parameters $b$ and $b_1$ are related to the radii squared of
corresponding $\eta(M)$-meson. Then the other constants ($C$, $C_1$,
$D_1$) are fixed by the normalization and orthogonality conditions:
\begin{equation}
\label{wfparam1}
\Psi^{(0)}_\eta\otimes\Psi^{(0)}_\eta=1, \quad
\Psi^{(1)}_\eta\otimes\Psi^{(1)}_\eta=1, \quad
\Psi^{(0)}_\eta\otimes\Psi^{(1)}_\eta=0.
\end{equation}
The convolution of the $\eta$-meson
 wave function at $q_\perp\ne 0$ determines
form factor of the $\eta$-meson,
$f^{(n)}_\eta(q^2_\perp)=
\left[\Psi^{(n)}_\eta\otimes\Psi^{(n)}_\eta\right]_{q_\perp\ne 0}$
thus allowing us to relate the parameter $b$ (or $b_1$)
at small $q^2_\perp$  to  $\eta$-meson radius
squared: $f_\eta(q^2_\perp)\simeq 1-\frac 16 R^2_\eta q^2_\perp$.
The $\eta$-meson form factor reads \cite{AMN}:
\begin{eqnarray}
f_\eta(q^2_\perp)&=&\frac{1}{16\pi^3}
\int\frac{dxd^2{k_\perp}}{x(1-x)^2}
\Psi^{(n)}_\eta(s)\Psi^{(n)}_\eta(s')
\left [\alpha(s+s'-q^2)+q^2 \right ],
\nonumber\\
\alpha&=&\frac{s+s'-q^2}{2(s+s') -\frac{(s'-s)^2}{q^2}-q^2}
\end{eqnarray}
where
$s'=(m^2+(\vec k_{\perp}-x\vec q_{\perp})^2)/x(1-x)$.
When working with a simple one-parameter wave function
representation of Eqs.
(\ref{wfparam}) and (\ref{wfparam1}), it is instructive to
compare the results with those obtained using  more
precise wave function patametrization; such a comparison can be
done for basic $2^1S_0q\bar q$ nonet.
The  $\eta$ and $\eta'$ wave functions (or those for its
$n\bar n$ and $s\bar s$
components) were found in \cite{AMN} basing on the data
for the transitions $\eta\to\gamma\gamma^*(Q^2)$,
$\eta\to\gamma\gamma^*(Q^2)$ at $Q^2\le 20$ GeV$^2$.
The calculated decay form factors
$F^{(0)}_{n\bar n\to\gamma\gamma}(k^2)$ and
$F^{(0)}_{s\bar s\to\gamma\gamma}(k^2)$ for these wave
functions are marked in Fig. 1b by rhombuses. The wave functions of
Ref. \cite{AMN} give the following mean radii squared for $n\bar n$
and $s\bar s$ components: $R^2_{n\bar n}=13.1$ GeV$^{-2}$ and
$R^2_{s\bar s}=11.7$ GeV$^{-2}$; in Fig. 1b we have drawn rhombuses for
these values of  radii. Solid curves in Fig. 1b represent
$F^{(0)}_{n\bar n\to\gamma\gamma}(0)$ and
$F^{(0)}_{s\bar s\to\gamma\gamma}(0)$ calculated by using the simple
exponential parametrization (\ref{wfparam}): we see that both
calculations coincide with each other within reasonable accuracy.
The coincidence of the results justifies
the exponential approximation for the calculation of
transition form facrors at $q_\perp^2 \sim 0$.

{\bf Results.}
The figure 2a demonstrates  calculation results for the
transition form factors
$n\bar n\to \gamma\gamma$ and $s\bar s\to \gamma\gamma$ when these
components refer to  $\eta$-mesons of the first
radial excitation multiplet. The form factor for the
$n\bar n$ component,  $F^{(1)}_{n\bar n\to\gamma\gamma}(0)$,
depends strongly on the mean radius squared,
increasing rapidly in the region $R^2_{n\bar n}\sim 14-24$ GeV$^{-2}$
(0.7-1.2 fm$^2$). As for $s\bar s$ component, the form factor
$F^{(1)}_{s\bar s\to\gamma\gamma}(0)$ is small; it changes sign at
$R^2_{s\bar s}\simeq 15$ GeV$^{-2}$. Therefore, one can neglect the
 contribution of the $s\bar s$ component into
$\gamma\gamma$ decay. Then
\begin{equation}
\Gamma_{\eta(M)\to\gamma\gamma}\simeq
\frac\pi 4\alpha^2M^3\cos^2{\phi}
F^{(1)\,2}_{n\bar n\to\gamma\gamma}=
\cos^2{\phi}\Gamma^{(1)}_{n\bar n\to\gamma\gamma}(0),
\end{equation}
where $\cos^2{\phi}$ is a probability for $n\bar n$ component in the
$\eta(M)$ meson. The calculated values
$\Gamma^{(1)}_{n\bar n\to\gamma\gamma}$ for $\eta(1295)$ and
$\eta(1440)$ are shown in Fig. 2b as functions of $R^2_{n\bar n}$.
A significant difference of widths
$\Gamma^{(1)}_{n\bar n\to\gamma\gamma}$ for $\eta(1295)$ and
$\eta(1440)$ is due to a strong dependence of partial width on
the $\eta$-meson mass, $
\Gamma_{\eta(M)\to\gamma\gamma}\sim M^3$.

{\bf Conclusion.}
We calculate the $\gamma\gamma$ partial width for $\eta(1295)$
and $\eta(1440)$ supposing these mesons to be members of the first
radial excitation nonet $2^1S_0q\bar q$. The calculation technique is
based on that developed in \cite{AMN} for the transition of mesons
from basic nonet $1^1S_0q\bar q$ into $\gamma^*(Q^2)\gamma$. The
$\gamma\gamma$ partial widths of $\eta(1295)$ and $\eta(1440)$ are
mainly determined by the flavour component
$n\bar n=(u\bar u+d\bar d)/\sqrt 2$ so
$\Gamma_{\eta(1295)\to\gamma\gamma}+\Gamma_{\eta(1440)\to\gamma\gamma}
 \simeq \Gamma^{(1)}_{n\bar n}$.
Partial widths strongly depend on the meson
radii squared:
$\Gamma_{\eta(1295)\to\gamma\gamma}+\Gamma_{\eta(1440)\to\gamma\gamma}
\simeq 0.04$ keV at $R^2_{\eta(X)}/R^2_\pi\leq 1.5$ and
$\Gamma_{\eta(1295)\to\gamma\gamma}+\Gamma_{\eta(1440)\to\gamma\gamma}
\simeq 0.2$ keV at $R^2_{\eta(X)}/R^2_\pi\simeq 2$.

The paper was partly supported by the RFBR grant 98-02-17236.


\begin{figure}[h]
\centerline{\epsfig{file=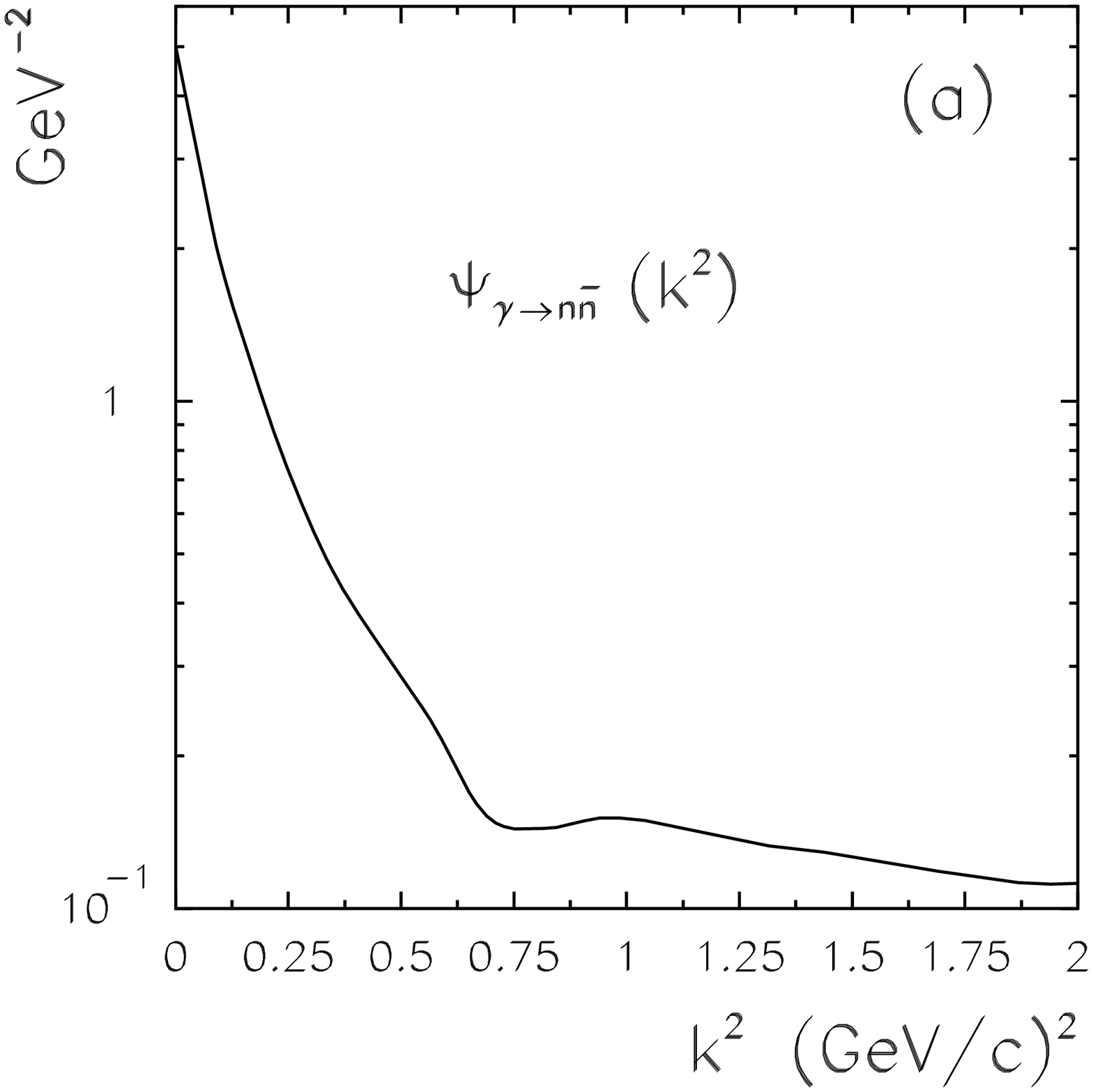,width=7.5cm}
            \epsfig{file=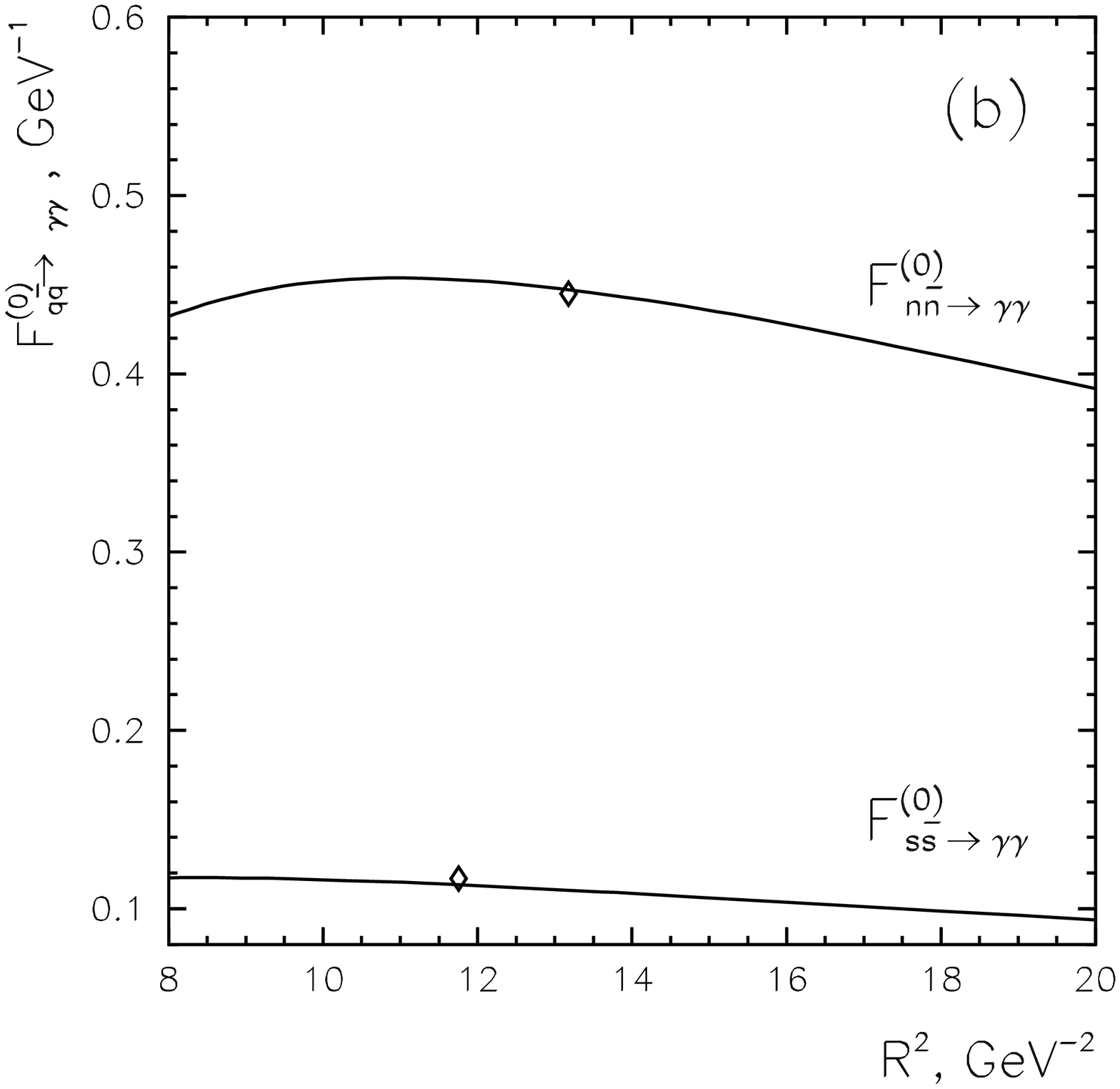,width=7.5cm}}
\caption{(a) Photon wave function
$\Psi_{\gamma\to n\bar n}(k^2)=g_\gamma(k^2)/(k^2+m^2)$, where
$k^2=$ $s/4-m^2$; the wave function for $s\bar s$ component is
obtained by replacement $m\to m_s$.
(b) Form factors $F^{(0)}_{n\bar n\to\gamma\gamma}(0)$ and
$F^{(0)}_{s\bar s\to\gamma\gamma}(0)$ as functions of mean radius
squared of $\eta$-meson. Solid curves present calculations with use
parametrization (5); the rhombuses give values calculated with use the
wave functions found in Ref. [3].}
\end{figure}

\begin{figure}
\centerline{\epsfig{file=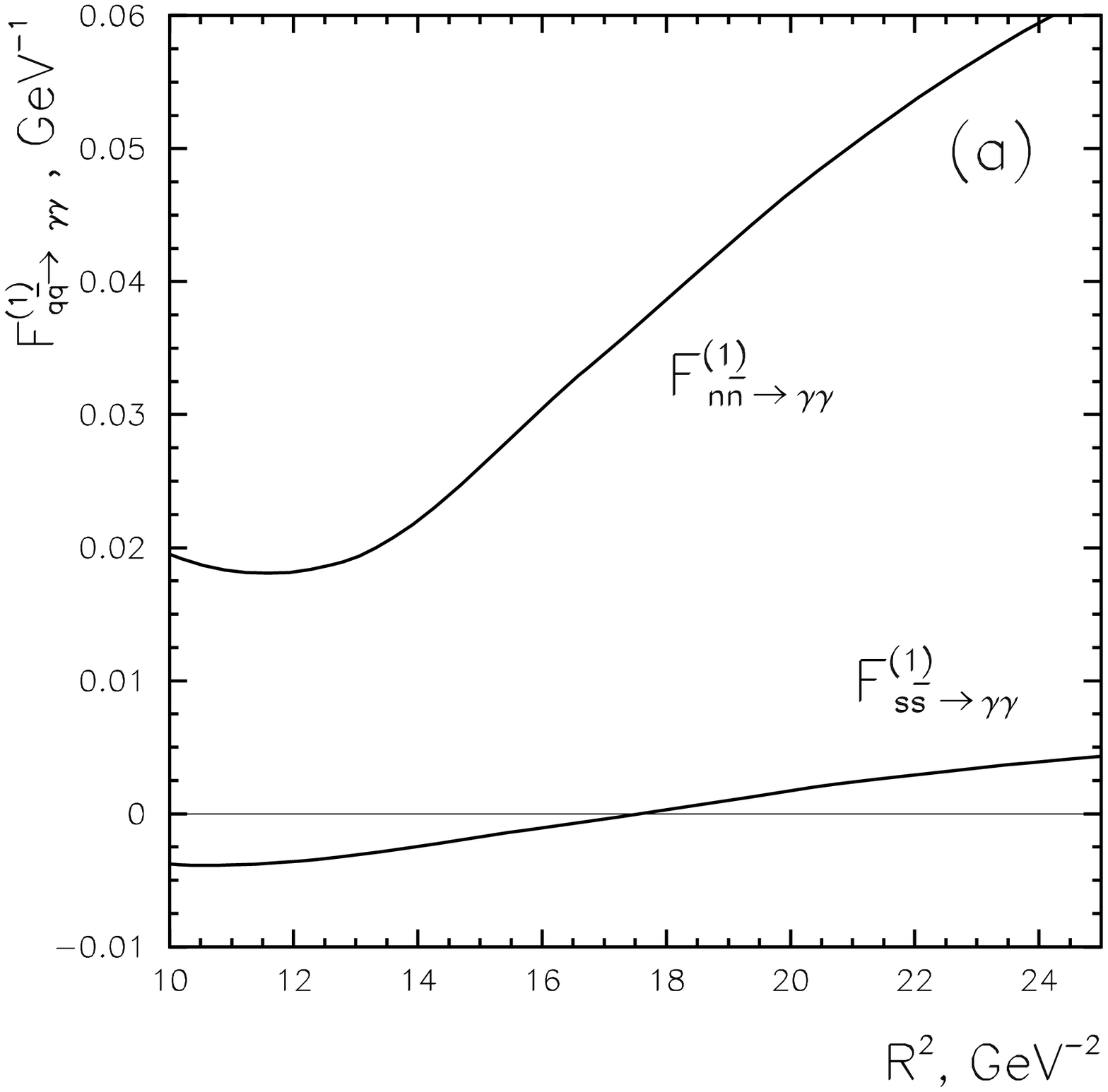,width=7.5cm}
            \epsfig{file=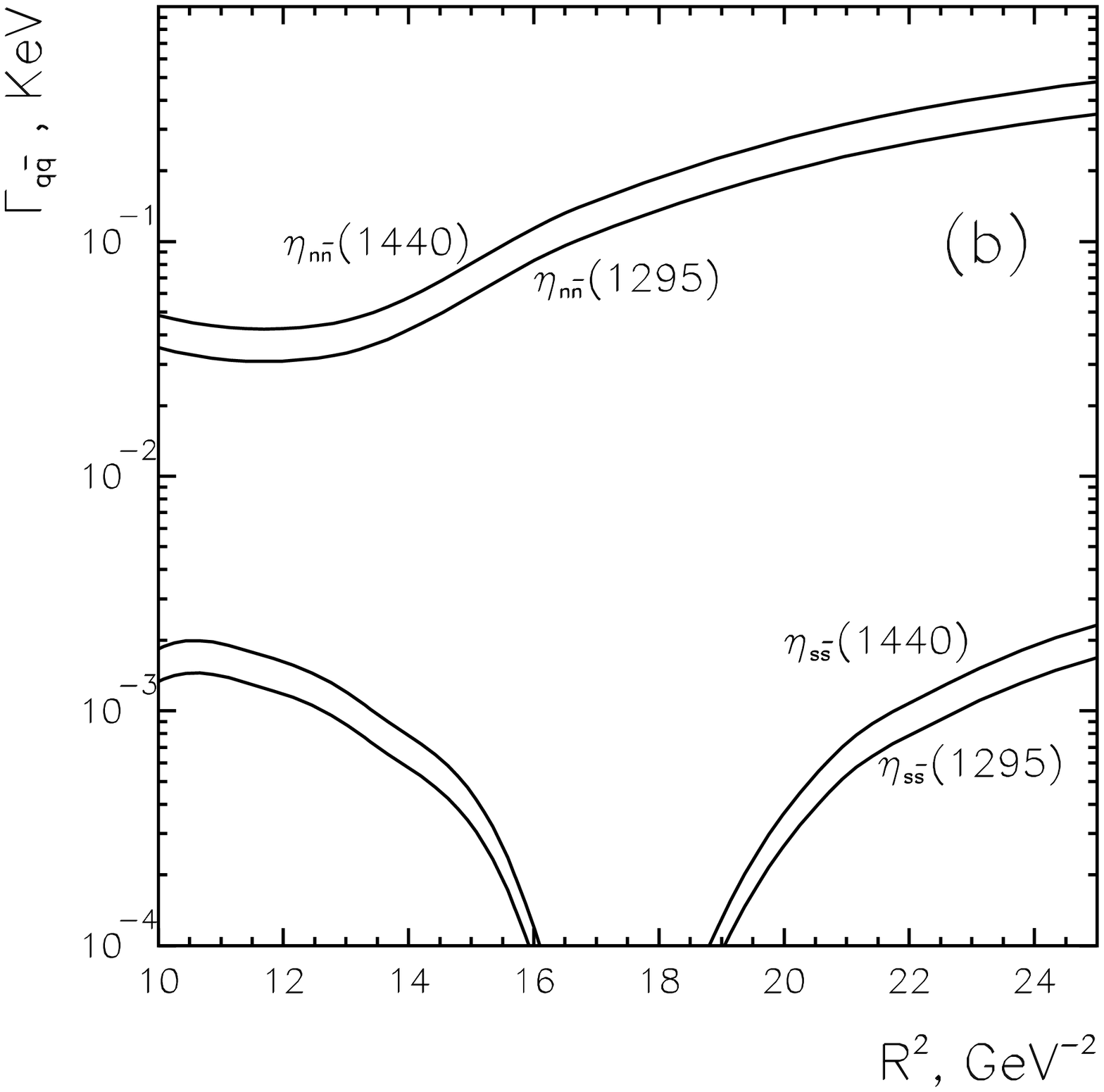,width=7.5cm}}
\caption{(a) The transition form factors
$F^{(1)}_{n\bar n\to\gamma\gamma}$ and
$F^{(1)}_{s\bar s\to\gamma\gamma}$ as functions of the
meson radius squared, $R^2$. (b)
The partial widths $\Gamma^{(1)}_{n\bar n\to\gamma\gamma}$ and
$\Gamma^{(1)}_{s\bar s\to\gamma\gamma}$ for
$\eta(1295)$ and $\eta(1440)$ as functions of $R^2$.}
\end{figure}

\end{document}